\documentstyle[11pt,aaspp4]{article}
\newcommand\beq{\begin{equation}}
\newcommand\eeq{\end{equation}}

\begin{document}

\title{Short GRBs and Mergers of Compact Objects: Observational Constraints}
\author{Rosalba Perna\altaffilmark{1,2,3} 
and Krzysztof Belczynski\altaffilmark{2,4,5}}

\altaffiltext{1}{Harvard Junior Fellow}

\altaffiltext{2}{Harvard-Smithsonian Center for Astrophysics, 60 Garden Street,
Cambridge, MA 02138}

\altaffiltext{3}{Istituto di Astrofisica Spaziale, C.N.R., 
via Fosso del Cavaliere, I-00133, Roma, Italy}

\altaffiltext{4}{Lindheimer Posdoctoral Fellow}

\altaffiltext{5}{Northwestern University, Dept. of Physics \& Astronomy,
 2145 Sheridan Rd., Evanston, IL 60208}

\begin{abstract}

GRB data accumulated over the years have shown that the distribution
of their time duration is bimodal.  While there is some evidence that
long bursts are associated with star-forming regions, nothing is
known regarding the class of short bursts. Their very short timescales
are hard to explain with the collapse of a massive star, but would be
naturally produced by the merger of two compact objects, such as two
neutron stars (NS), or a neutron star and a black hole (BH).  As for
the case of long bursts, afterglow obervations for short bursts should
help reveal their origin. By using updated population synthesis code
calculations, we simulate a cosmological population of merging NS-NS
and NS-BH, and compute the distribution of their galactic off-sets, the
density distribution of their environment, and, if indeed associated
with GRBs, their expected afterglow characteristics.

\end{abstract}


\section{Introduction}

Since data on GRBs started to accumulate over the past two decades, it
was recognized that their time distribution appears to be
bimodal\footnote{Note however the recent claims about the possibility
that a third class may exist (Balastegui et al. 2001).}, with about
25\% of bursts having a short duration, of mean $\sim 0.2$ sec, and
the rest having a much longer duration, of mean $\sim 20$ s (Mazets et
al. 1981; Hurley et al. 1992; Kouveliotou et al. 1993; Norris et
al. 2000).  The separation between the two classes appears to be around
2 s.  While the two classes of bursts seem to have similar (isotropic)
spatial distributions, they differ in several other respects. Short
bursts tend to have harder spectra than long bursts (Kouveliotou et
al. 1993; Dezelay et al. 1996), and about 20 times less
fluence\footnote{Lee \& Petrosian (1997) showed indeed that there is a
highly significant positive correlation between the burst fluence and
duration.} and it has also been presented some evidence that their
number-intensity distribution (Belli 1997; Tavani 1998) differs from
that of the longer class.  A recent analysis by Norris et al. (2000)
of the temporal properties of the bursts, such as the distribution of
number of pulses per burst, pulse width, and intervals between pulses,
clearly slowed that the two classes of long and short bursts are
disjoint.

The existence of two distinct populations of GRBs might very well be
an indication of the presence of two distinct types of progenitors.
The currently favoured GRB models can be divided into two classes:
models involving mergers of two compact objects (Goodman 1986; Eichler
et al. 1989; Paczynski 1991; Narayan et al. 1992; Meszaros \& Rees
1992; Katz \& Canel 1992), and models involving the collapse of a
massive star (Woosley 1993; Paczynski 1998; MacFadyen \& Woosley 1999;
Vietri \& Stella 1998).  According to the internal shock model (see
e.g. Piran 1999 for a review) for the production of the observed
$\gamma$-ray emission, the duration of the event, whatever it is, is
very likely to be a direct measure of the time interval during which
the powering engine is active.  Simulations of mergers of two compact
objects have shown that the duration of the neutrino-driven wind
possibly producing the GRB (Ruffert \& Janka 1999) is less than a
second. On the other hand, the relativisitc outflow generated by the
collapse of a massive star (MacFadyen \& Woosley 1999) can last
several tens of seconds.  Therefore, if one were to associate the two
classes of GRBs with two classes of models, it would be natural to
associate long GRBs with the collapse of massive stars and short ones
with mergers of two compact objects.

Traditionally, it is considered that an important way to test the
above assumption and possibly distinguish between the two classes of
models is by determining the location in which the bursts occur.
Massive stars have short lifetimes, and therefore are expected to die
close to where they are born, that is in dense and dusty
environments. On the other hand, compact objects receive kicks when
they are born, and are therefore expected to travel far from their
birthplaces.

Along these lines, several methods have been proposed to constrain the
GRB location and the characteristics of their environment.  The
long-lived remnants resulting from the interaction between GRBs and
their afterglows with the surrounding medium can be identified in
nearby galaxies based on their spectral signatures (Wang 1999; Perna,
Raymond \& Loeb 2000; Perna \& Raymond 2000) or their dynamical
interaction with the medium (Efremov, Elmegreen \& Hodge 1998; Loeb \&
Perna 1998, Ayal \& Piran 2001), therefore allowing a close study of
the GRB sites.  On the much shorter time scale during which the
afterglow propagates in the medium, several independent analyses can
be made.  When multifrequency data are available, then a determination
of the various break frequencies and the peak flux in the afterglow spectrum (Sari,
Piran \& Narayan 1998) allows to constrain the burst parameters; this
has been done in several cases by a number of authors (e.g. Wijers \& Galama 1999; 
Panaitescu \& Kumar 2001a).  Time
variability of absorption lines due to the gradual photoionization of
the medium by the X-ray UV radiation is also sensitive to the type of
environment (Perna \& Loeb 1998; B\"ottcher et al. 1998; Lazzati et
al. 2001), as it is the time delay between the $\gamma$-ray emission
and the onset of the afterglow (Vietri 2000).

So far, afterglow observations have only been possible for long 
GRBs\footnote{See however Lazzati, Ramirez-Ruiz and Ghisellini 2001
for a possible detection in the BATSE data.}.
For this class of bursts, there has been mounting evidence that they
are associated with the collapse of massive stars. 
Bloom, Kulkarni \& Djorgovski (2001) compared
the observed offset distribution of 20 GRBs with the theoretical
predictions of two models, one which is representative of collapsars
and promptly bursting binaries (such as binaries in which the black
hole merges with the helium core of an evolved star during a common
envelope phase), and another representative of delayed merging
remnants.  They found that the latter population can be ruled out to a
high confidence level. Their conclusions are strengthened by the
observed correlation of GRB locations with the UV light of their
hosts, which is strongly suggestive of the occurrence of GRBs in
star-forming regions (see also Sahu et al. 1997, Kulkarni et al. 1998,
Fruchter et al. 1999, Kulkarni et al. 1999). 

An analysis of the
evolution of the X-ray prompt emission of GRB 980329 and GRB 780506 by
Lazzati \& Perna (2001) has led to similar conclusions. An intriguing
hint towards the connection of GRBs with the collapse of massive stars
has been provided by the presence of a bump, interpreted as an
underlying supernova component, in the light curve of GRB 900326
(Bloom et al. 1999) and GRB 970228 (Reichart 1999, Galama et
al. 2000). The evidence for extinction by dust of some burst
afterglows has provided further support to the GRB-collapsar
connection. Finally, the recent detection of an iron line in the
afterglow spectrum of 5 GRBs (Piro et al. 1999, 2000; Yoshida et
al. 1999; Antonelli et al. 2000; Amati et al. 2000) provides evidence
for the presence of dense matter in the vicinity of the burst sites
(Vietri et al. 1999; Weth et al. 2000; Lazzati, Covino \& Ghisellini
2000a).  The association between GRBs and collapsars, once well
established, would have very important implications for our
understanding of the star formation history in the universe (Blain \&
Natarajan 1999).

This huge wealth of information on the class of long GRBs has been
gathered as a result of the afterglow observations.  No similar types
of studies have been possible for the short bursts so far.  At this
point, optical and radio searches have been performed only for the 4
bursts that were well pin pointed by the Interplanetary Network
(Hurley et al. 2001), but the search did not lead to detections.  The
situation is however going to change with HETE II and then {\em
Swift}, which will provide quick arc-minute localizations;
longer-wavelength follow-ups will then be possible also for this class
of short events, therefore allowing to do the same type of science for
them as well.

As already mentioned, a strong candidate for the class of short bursts
is provided by the coalescence of two compact objects, whose
timescales and energetics are compatible with those inferred for the
class of short GRBs.  Therefore an analysis of the afterglow
properties that such a population would have is needed. This is the
goal of this work. More specifically, in this paper we use updated
population synthesis code calculations to study the afterglows that a
cosmological population of GRBs due to mergers of compact objects
should have. We consider
two classes of progenitors: double neutron stars, and neutron star --
black hole. The former population includes the new class of short-lived
neutron stars identified by Belczynski \& Kalogera (2001), 
Belczynski, Kalogera \& Bulik (2001c) and Belczynski, Bulik \& Kalogera,
(2001a).
As it will be discussed in the following, this population has a very 
short lifetime and it dominates the merger rates. The environment of this
population would therefore be more similar to that of collapsars and 
helium star-black hole mergers. On the other hand, the population
of NS-BH binaries has a much longer lifetime, and therefore it has
time to travel further away from its birthsite. 
More generally, we want to point out that, whereas the motivation
of this paper has been a detailed study of merger events in relation
to the class of short bursts, however our results regarding the
new class of tight NS-NS binaries can be potentially relevant also
for the class of long GRBs. In fact, as discussed above, some of the
evidence on the connection of long GRBs with massive stars is based
on the association of long GRBs with star forming regions, but this
would also be the case for the new population of tight NS-NS binaries.

To generate our simulated set of data, we incorporate the results of
the population synthesis code {\em StarTrack} by Belczynski et al.\
(2001c) within the context of a cosmological model which, with the help
of a Monte Carlo type approach, accounts for (i) the redshift
distribution of the merger events, (ii) the mass distribution of the
galaxies where the events occur (which is a function of redshift)
using a Press-Schechter type formalism (Press-Schechter 1974); (iii) the
redshift dependence of the probability that a certain merger occurs at
a given position within a galaxy of a given mass. This last effect
is particularly important for the population of NS-BH binaries, whose
lifetime can be comparable with the Hubble time, and it has not been
considered so far. A byproduct of our computation is the distribution
of the offsets that merging binaries have from the centers of their
host galaxies.  For each merger event, the density of the surrounding
medium is also determined within the simulation itself.  The other
parameters that are needed to compute the afterglow intensity for each
event are randomly drawn from distributions which have the typical
values found in the afterglow modelling of long GRBs.

The paper is organized as follows: in \S 2 we describe the various 
elements of the computation, which include the population synthesis code,
the new population of double NS-NS, the galaxy potential
and its density profile, and the afterglow modelling. The results
of the simulation of the data are presented in \S 3, while \S 4 is devoted to
a discussion with conclusions. 

\section{Model}

In this section we describe the various ingredients of the calculation.

\subsection{Population synthesis model}

We use the {\em StarTrack} population synthesis code 
developed by Belczynski et al. (2001c).
In the following we summarize the basic assumptions and ideas 
of the code.
However, for more detailed description of the {\em StarTrack} we 
refer the reader to Belczynski et al.\ (2001c).

The evolution of single stars is based on the analytic formulae 
derived by Hurley, Pols \& Tout (2000). 
With these formulae we are able to calculate the evolution of 
stars for Zero Age Main Sequence (ZAMS) masses 0.5--$100\,
{\rm M}_\odot$ and for metallicities:  $Z=0.0001 - 0.03$.

Stellar evolution is followed from ZAMS through different 
evolutionary phases depending on the initial (ZAMS) stellar mass: 
Main Sequence, Hertzsprung Gap, Red Giant Branch, 
Core Helium Burning, Asymptotic Giant Branch, 
and for stars stripped off their hydrogen-rich layers: 
Helium Main Sequence, Helium Giant Branch.  
We end the evolutionary calculations at the formation of a stellar
remnant:  a white dwarf, a neutron star or a black hole.  

There are two modifications to the original Hurley et al. 
(2000) formulae concerning the treatment of (i) final remnant masses 
(see Belczynski et al.\ 2001c), and (ii) Helium-star evolution (see 
Belczynski \& Kalogera 2001).
 
The {\em StarTrack} code employs Monte Carlo techniques to model the evolutionary 
history and coalescence rates of binary compact objects, e.g., NS-NS and NS-BH.  
A binary system is described by four initial parameters: the mass $M_1$ of the 
primary (the component which is initially more massive), 
the mass ratio $q$ between the secondary and the primary, the semi-major axis of the
orbit $A$, and the orbital eccentricity $e$. 
Each of these initial parameters is drawn from a distribution. More specifically,
the mass of the primary is drawn from the Scalo initial mass function (Scalo 1986),
\beq
\Psi(M_1)\propto M_1^{-2.7}
\label{eq:Mf}
\eeq
and within the mass range $M_1=5-100M_\odot$. The distribution of the mass ratios is taken
to be
\beq
\Phi(q)=1,\;\;\;\;\; 0\leq q\leq 1\,,
\label{fq} 
\eeq
following Bethe \& Brown (1998). The initial binary separations, $A$, are assumed as in
Abt (1993)
\beq
\Gamma(A)\propto\frac{1}{A}\;,
\label{eq:fA}
\eeq
and finally, the initial distribution of the binary eccentricity is taken
following Duquennoy \& Mayor (1991).
\beq
\eta(e)=2e
\label{eq:fe}\;,\;\;\;\;\;0\leq e\leq 1\;.
\eeq

Further, the initial distribution of binaries is assumed to follow the mass
distribution in the young disk (Paczynski 1990), i.e.
\beq
P_{\rm bin}(R,z)dRdz=P(R)dRp(z)dz\propto R \,e^{-R/R_0}e^{-z/z_0}dR\,dz\;.
\label{eq:fP}
\eeq 
For a galaxy like the Milky Way, $R_0=4.5$ kpc up to $R_{\rm max}=20$ kpc, and
$z_0=75$ pc. These parameters are assumed to scale with the galaxy mass as
discussed in the following section. 

As we are interested only in NS-NS and NS-BH systems, we evolve only massive
binaries, with primaries more massive than 5 M$_\odot$.  We generate a
large number ($N = 1.6 \times 10^7$) of primordial binaries and evolve them
until formation of the remnant system.  During the evolution of every
system we take into account the effects of wind mass-loss, asymmetric
supernova (SN) explosions, binary interactions (conservative/non-conservative mass
transfers, common envelope phases) on the binary orbit and the
binary components. 
We also include effects of accretion onto compact objects in 
common envelope (CE) phases 
(Brown 1995; Bethe and Brown 1998) and rejuvenation of binary components
during mass transfer episodes.
Once a binary consists of two compact remnants (NS
or BH), we calculate its merger lifetime, the time until the
components merge due to gravitational radiation and associated orbital
decay.  

The {\em StarTrack} code was used in its standard mode, described by
the set of parameters which are thought to represent at best our understanding of
stellar single and binary evolution.
 (1) {\em Kick velocities.} Compact objects receive natal kicks during 
supernova explosions, when they are formed.
Neutron star kicks are drawn from a weighted sum of two Maxwellian
distributions with $\sigma=175$\,km\,s$^{-1}$ (80\%) and
$\sigma=700$\,km\,s$^{-1}$ (20\%) (Cordes \& Chernoff 1997).
For black holes formed via partial fall back we use smaller kicks, 
but drawn from the same distribution as for NS. 
The kick scales with the amount of material ejected in SN explosion 
or inversely with the amount of falling back material 
(the bigger the fall back,the smaller the kick).
And for BHs formed in direct collapse of massive stars we do not apply 
any kicks, as in those cases no supernova explosion accompanies the formation of 
such objects. 
 (2) {\em Maximum NS mass.} We adopt a conservative value of $M_{\rm
max}=3$\,M$_\odot$ (e.g., Kalogera \& Baym 1996). It affects the relative
fractions of NS and black holes and the outcome of NS hyper--critical
accretion in CE phases;
 (3) {\em Common envelope efficiency.} We assume $\alpha_{\rm
CE}\times\lambda = 1.0$, where $\alpha_{\rm CE}$ is the efficiency with which
orbital energy is used to unbind the stellar envelope, and $\lambda$ is a
measure of the central concentration of the giant;
 (4) {\em Non--conservative mass transfer.} In cases of dynamically stable
mass transfer between non--degenerate stars, we allow for mass and angular
momentum loss from the binary (see Podsiadlowski, Joss, \& Hsu 1992),
assuming that the fraction $f_{\rm a}$ of the mass lost from the donor is
accreted to the companion, and the rest ($1-f_{\rm a}$) is lost from the
system with specific angular momentum equal to $2\pi jA^2$/P,
where $A$ is the orbital separation and P the period.
We adopt $f_{\rm a}=0.5$ and $j=1$.
 (5) {\em Star formation history.} 
We assume that star formation has been continuous in the disk of a given 
galaxy.
We start the evolution of a single or a binary system $t_{\rm birth}$\ ago, 
and follow it to the t(z), where t(z) is the present time at a given
redshift $z$ of the galaxy.
The birth time $t_{\rm birth}$ is drawn randomly within the range
0--t(z), which corresponds to continuous star formation rate.
(6) {\em Initial Binarity}
We assume a binary fraction of $f_{\rm bi}=0.5$, which means that for 
any 150 stars we evolve, we have 50 binary systems and 50 single stars.
(7) {\em Metallicity}
We assume solar metallicity $Z=0.02$.
(8) {\em Stellar Winds}
The single-star models we use (Hurley et al. 2000) include the effects 
of mass loss due to stellar winds.
Mass loss rates are adopted from the literature for different 
evolutionary phases, that is 
for H-rich stars on MS (Nieuwenhuijzen \& de Jager (1990), 
for RG branch stars (Kudritzki \& Reimers 1978) using the $Z$
dependence of Kudritzki et al. (1989), for AGB (Vassiliadis \& Wood 1993), and
finally for Luminous Blue Variables (Hurley et al.\ 2000).
For He-rich stars W--R mass loss is included using rates derived by
Hamann, Koesterke \& Wessolowski (1995) and modified by Hurley et al. 
(2000).

The population synthesis code allows us to compute the probability
distribution, $P_{\rm merg}(t)$, of a merger of a given type as a
function of the time $t$ since the formation of the system.  It also
yields the probability distribution, $P_{\rm loc}(R,\eta;t,M_{\rm
gal})$, that a merger in a galaxy of mass $M_{\rm gal}$ after a time
$t$ of formation of the system occurs at the position $(R,\eta)$ 
from the galaxy center. 

We have considered a grid of galaxy masses with fifteen equal
logarithmically-spaced intervals in the range
$\{10^8-10^{11}\}M_\odot$. For each value of the mass, ten values of
the cosmological time corresponding to redshifts linearly spaced
in the interval \{0,10\} were considered. For each of the values of
$M_{\rm gal}$ and $z$ on the grid, the population synthesis code was
run to obtain probability distributions for the positions \{$R(M_{\rm
gal},z),\eta(M_{\rm gal},z)$\} of the mergers taking place in the galaxy
of mass $M_{\rm gal}$ at the redshift $z$.  Probability distributions for
values of $M_{\rm gal}$ and $z$ not on the grid were obtained by
interpolation.

\subsection{Double neutron star binaries}

Belczynski \& Kalogera (2001) and Belczynski et al. (2001a, 2001c) identified 
new subpopulations of NS-NS binaries.
The new subpopulations dominate the group of coalescing NS-NS systems, and
due to their unique characteristics, they predominantly merge inside the
host galaxies. Given the importance of these populations to our conclusions,
in what follows we briefly summarize the results of Belczynski \& Kalogera
(2001) and Belczynski et al. (2001a, 2001c).

Double neutron stars are formed in various ways, including more than
14 different evolutionary channels, as discussed by Belczynski et al.\
(2001c).  However, the whole population of coalescing NS-NS systems can
be divided into 3 subgroups.

{\em Group I} consists of non-recycled NS-NS systems, which terminate
their evolution in a double CE of two helium giants.  Two bare
Carbon-Oxygen (CO) cores emerge after envelope ejection, and they form
neutron stars in two consecutive SN type Ic explosions.  Provided that
the system is not disrupted by SN kicks and mass loss, the two NS form
a tight binary, with the unique characteristic that none of NS had a
chance to be recycled.  For more details see Belczynski \& Kalogera
(2001).  {\em Group II} includes all the systems which finished their
evolution through single CE phase, with a helium giant donor and a NS
companion.  During the CE phase, the neutron star accretes material
from the giant envelope, becoming most probably a recycled pulsar.
The CO core of the Helium giant, soon after CE phase is finished,
forms another neutron star.  The system has a good chance to survive even a high
kick that the newly born NS may recieve, because after the CE episode it is very
tight and well bound.  For more details see Belczynski et al. (2001a).
{\em Group III} consists of all the other NS-NS systems formed,
through more or less classical channels (Bhattacharya \& van den
Heuvel 1991).

Group II strongly dominates the population of coalescing NS-NS systems (81\%)
over group III (11\%) and I (8\%).
This is due to the fact that we allow for helium star radial evolution, and
usually just prior to the formation of tight (coalescing) NS-NS system we
encounter one extra CE episode, as compared to classical channels.
This has major consequences for the merger time distribution of the NS-NS
population, and in turn for the distribution of NS-NS merger sites around their
host galaxies.
Merger times of classical systems are comparable with Hubble time, and that 
gives them ample time to escape from their host galaxies.
As it has been shown in previous studies (e.g., Bulik, Belczynski \&
Zbijewski 1999; Bloom, Pols \& Sigurdsson 1999) which did not include Helium 
star detailed radial evolution, a significant fraction of the NS-NS population tended to 
merge outside host galaxies, exactly like our group of classical systems. 
In contrast, the binaries of Group I and II, due to the extra CE episode, are
tighter, and their merger times are much shorter: of order of $\sim 1$\
Myr. 
Thus even if they acquire high systematic velocities, due to the asymetric SN
explosions, they will merge within the host galaxies, near the places they
were born. 
Since Group I and II dominate the population, the overall NS-NS
distribution of merger sites will follow the distribution of primordial
binaries or star formation regions in the host galaxy.

Note that NS-NS binary systems with rather short lifetimes had also
been proposed by Tutukov \& Youngelson (1993, 1994), and discussed in the
context of GRBs as well. They find merger times which are generally
shorter than those found in other studies (e.g.  Portegies-Zwart \&
Youngelson 1998; Fryer et al. 1999), but not as short as those found
by Belczynski et al. (2001a, 2001c). 
In the Tutukov \& Youngelson  scenario the short time scales are mainly 
the result of assumption that the secondary star, once it becomes a low 
mass helium rich star, can initiate an extra MT phase.
In the Belczynski et al. scenario, the short time scales are the
result of allowing, both the primary and the secondary star, to initiate 
an extra MT or CE phase.   
Moreover, in contrast to Tutukov \& Youngelson, Belczynski et al. scenario 
incorporates the assumption that NS do receive natal kicks.
Due to the natal kicks, the widest binaries are disrupted, and the 
surviving ones gain high eccentricities, which further reduces their 
merger times (i.e., lifetimes).   
We want to stress that the Belczynski et al. (2001a, 2001c) results rely on 
the assumption that low-mass helium stars can actually survive the CE phase, 
and this assumption needs yet to be tested through detailed 
hydrodynamical calculations.

The distribution of merger times of BH-NS systems is similar to that of the
classical NS-NS subpopulation (long merger times, $\ga 0.1-1$ Gyr).
However, these systems have obtained smaller systematic velocities as 
on average black holes receive smaller kicks than neutron stars.
As a  result, a certain fraction of BH-NS binaries escape and merge outside of 
the host galaxies; however,  the fraction of escaping BH-NS systems is
smaller than that of classical NS-NS binaries (this is a result seen in 
previous studies, in which only classical NS-NS were considered). 
More detailed study of all proposed binary GRB progenitors can be found 
in Belczynski, Bulik \& Rudak (2001b). 

\subsection{Cosmic event rate}

The merger rate of a given type of progenitors is obtained by combining the results of the
population sunthesis code with the star formation history. 
Let $R_{sfr}(t)$ be the cosmic star-formation rate (SFR) at time $t$;
here we adopt the SFR of Rowan-Robinson (1999).  
Let $f_i$ be the mass fraction (of all stars, single and binary, in mass 
range [0.08-100 $M_\odot$])  
leading to formation of the GRB progenitor of type ``$i$''. 
The cosmic event rate for that type of mergers at redshift $z$ is then given by 
\beq
R_{\rm merg}(z)= \int_{t(z)}^{t(z=\infty)}R_{sfr}(t')f_i\,P_{\rm merg}(t(z)-t')\,dt' \;, 
\label{eq:Rmerg}
\eeq
where $P_{\rm merg}(t)$ is defined in the previous section, and
$dt={\rm H}_0^{-1}dz(1+z)^{-1}[(1+\Omega_m z)(1+z)^2-z(z+2)\Omega_\Lambda]^{-1/2}$.
The rate of events up to redshift $z$ for the GRB progenitor of type
``$i$'' is correspondingly given by
\beq
R_i(<z)=4\pi\int_0^z r^2_{z'}\frac{dr_{z'}}{dz'}\frac{R_{\rm
merg}(z')}{1+z'}dz'\;,
\label{eq:Ri}
\eeq
with $r_z=cH_0^{-1}\int_0^z[\Omega_m(1+z')^3+\Omega_\Lambda]^{-1/2}dz'$.

Throughout the paper we adopt a flat cosmology  with h=0.65, density parameter
$\Omega_m=0.3$ and cosmological constant $\Omega_\Lambda=0.7$.

\subsection{Galaxy model}

The potential of a spiral galaxy can be described as the sum of three
components: a bulge, a disk and a dark matter halo. A good approximation
to the potential of the disk and the bulge has been proposed by
Miyamoto \& Nagai (1975):
\beq
\Phi_{b,d}(R,\eta)=\frac{GM_{d,b}}{\sqrt{R^2+(a_{b,d}+\sqrt{z^2+c_{b,d}^2})^2}}\;, 
\label{eq:potbd}
\eeq 
where $a_{b,d}$ and $c_{b,d}$ are parameters (which depend on whether one
considers the bulge or the disk), $M_{d,b}$ is the mass either of the 
bulge or the disk, $R=\sqrt{x^2+y^2}$ is the coordinate in the
plane of the disk, and $\eta$ is the coordinate in the plane
perpendicular to the disk.  

The mass density distribution associated with the potential $\Phi_{b,d}(R,\eta)$ is
\beq
\rho_{d.b}(R,\eta)=\left(\frac{c_{b,d}^2 M_{b,d}}{4\pi}\right)
\frac{a_{b,d}^2 R^2 + (a_{b,d}+3\sqrt{\eta^2+c_{b,d}^2})(a_{b,d}
+\sqrt{\eta^2+c_{b,d}^2})^2}
{[R^2 + (a_{b,d}+\sqrt{\eta^2+c_{b,d}^2})^2]^{5/2}
(\eta^2+c_{b,d}^2)^{3/2}}\;.
\label{eq:rhobd}
\eeq
The dark matter halo is spherically symmetric, and described by the
potential
\beq
\Phi(r)=-\frac{GM_h}{r_0}\left[\frac{1}{2}\ln\left(1+\frac{r^2}{r_0^2}\right)
+\frac{r_0}{r}\arctan\left(\frac{r}{r_0}\right)\right]\;,
\label{eq:poth}
\eeq
where $r=\sqrt{R^2+\eta^2}$, and $r_0$ is a parameter.
The corresponding mass density distribution is 
\beq
\rho_h(r)=\frac{\rho_0}{1+(r/r_0)^2}\;,
\label{eq:rhoh}
\eeq 
where $\rho_0\equiv M_h/(4\pi r_0^3)$. The fraction of mass in gas is assumed to be
$f_{\rm gas}=0.5$ for the bulge and the disk, and $f_{\rm gas}=
\Omega_b/\Omega=0.04$ (Bahcall et al. 1999) for the halo.   
For the Milky Way, $a_b= 0$ kpc, $c_b=0.277$ kpc, $a_d=4.2$ kpc,
$c_d=0.198$ kpc, $M_b=1.12\times 10^{10}M_\odot$, $M_d=8.78\times
10^{10}M_\odot$, $r_0=6.0$ kpc, and $M_h=5.0\times 10^{10}M_\odot$.
$N$-body simulations by Bullock et al (2001) have shown that the
redshift evolution of the core radius of a halo, $r_0$, is roughly
constant, and this is what we assume here. Finally, we assume that the ratio
of the parameters describing the various components of the galaxy is
constant, independent of the galaxy mass\footnote{We 
neglect the scatter in the the ratio
between the disk size and the virial radius of the halo
and assume a typical value for it, calibrated on that of the Milky Way.
Introducing this scatter here would significantly increase our
computation time (requiring a much larger number of runs
of the population synthesis code), but would not affect our main
results.}, and that they scale with the
galaxy mass as $M^{1/2}$ (constant surface brightness; see e.g. Binney \& Tremaine 1994).

At redshift $z$, the probability distribution $P_{\rm gal}(M,z)$ of
having a merger in a galaxy of mass $M$ can be approximated by the
Press-Schechter function (yielding the probability of finding a
halo with mass $M$ at redshift $z$) convolved with the mean number of
galaxies, $N_{\rm gal}$, per halo of mass $M$.
For the latter, we use
the analytical approximation\footnote{This relation tries to capture
two important physical effects, that is the fact that, at large masses,
the gas cooling time becomes larger than the Hubble time, hence suppressing
galaxy formation in large mass halos; in small-mass halos, on the other hand,
phenomena such as supernova winds can blow away the gas from halos, also
suppressing galaxy formation, and this leads to a cutoff at small halo masses.} 
derived by Scoccimarro et al. (2001),
$\left <N_{\rm gal}\right >=\left < N_B \right> + \left < N_R \right >$, 
where $N_B$ and $N_R$ represent the
number of blue and red galaxies, respectively, per halo of mass $M$,
and their mean is given by $\left < N_B \right
>=0.7(M/M_B)^{\alpha_B}$ and $\left < N_R \right
>=0.7(M/M_R)^{\alpha_R}$, respectively.  The fit parameters
are $\alpha_B=0$ for $10^{11}M_\odot h^{-1}\le M\le M_B$, $\alpha_B=0.8$
for $M>M_B$, $\alpha_R=0.9$, and $M_R=2.5\times 10^{12} M_\odot h$.

The merger rate of compact objects as a function of galaxy mass
is not a well-constrained quantity. The simplest assumption is
that it simply scales with the mass of the galaxy. We adopt this
model as our ``standard'' model, but we will also explore how
the results  change if more weight (than the simple rescaling
with mass) is given to galaxies with smaller mass, given the observation
(Babul \& Ferguson 1996) that small mass galaxies might have an increased 
star formation rate. We parameterize the weight of the rate on the galaxy
mass though the quantity $M^\beta$, so that the probability of finding
a galaxy of mass $M$ at redshift $z$ is given by   
\beq
P_{\rm merg}(M,z)dM=A P_{\rm gal}(M,z)M^\beta dM\;,
\label{eq:pgal}
\eeq
where $A$ is a normalization factor so that $\int dM P_{\rm merg}(M,z)=1.$
We consider the values $\beta=1$ and $\beta=0.5$.    

Figures 1a, 1b and 1c show the probability distribution of projected
distances $P_{\rm loc}(R,\eta; z,M_{\rm gal})$ for a face-on galaxy
(i.e. $d_{\rm proj}=R$) at the redshifts $z=0,3,6$ and for a wide
range in galaxy masses, $M=10^8 M_\odot$ in Fig.1a, $M=6\times 10^{9}
M_\odot$ in Fig.1b, and $M=10^{11} M_\odot$ in Fig.1c.  While the
distribution of NS-NS mergers does not evolve much with redshift (due
to its very short lifetime), the population of NS-BH mergers does
evolve significantly, especially in small-mass galaxies\footnote{For
the class of long GRBs, their association with blue, starbust galaxies
(in the great majority of the cases where an host could be identified)
suggest that their masses lie in the range $10^8-10^{10}M_\odot$ (see
Bloom et al.  2001).}.  Therefore it is important that, for a merger
occurring at redshift $z$, the location within the host is determined
according to the probability distribution at that particular redshift.

\subsection{GRBs and afterglow parameters}

Even though the ultimate energy source that powers short bursts could
be different from that associated with long GRBs, there is no reason to expect
that the  physical mechanisms 
that produce the gamma radiation and the afterglow should also be different.
In the standard fireball scenario (see e.g. Piran 1999
for a review), GRBs are generated by internal 
shocks in the expanding fireball, while the afterglow is produced by 
shocks created by the interaction between the relativistically expanding matter
of the fireball itself and the surrounding  medium (the so-called ``external'' shocks). 
Moreover, no correlation is expected between the duration of the burst and
the decay rate of its afterglow. Therefore, to predict the afterglow 
properties of the population of short bursts, we use the theory developed
and used to study long bursts (e.g. Sari 1997; Waxman 1997; Sari, Narayan \& Piran 1998). 
In the standard model, GRB afterglows
are thought to be the result of synchrotron emission by Fermi-accelerated
electrons behind the expanding shock. The electrons have a power-law distribution
of Lorentz factors above a minimum value $\gamma_m$.     
As noted by Sari et al. (whose formalism we adopt here), 
there is also a threshold Lorentz
factor, $\gamma_c$, above which electrons rapidly loose their energy
to radiation, and  cool down to a Lorentz factor  $\sim \gamma_c$. 
Under this condition (denominated {\em fast cooling regime}),
the flux at the observation frequency $\nu$ is
\beq
 F_\nu  = F_{\nu,{\rm max}} \left\{
  \begin{array}{ll}
     (\nu/\nu_c)^{1/3}, & \hbox{$ \nu < \nu_c$} \\
     (\nu/\nu_c)^{-1/2}, & \hbox{$ \nu_c\le \nu < \nu_m$} \\
     (\nu_m/\nu_c)^{-1/2}(\nu/\nu_m)^{-p/2}, & \hbox{$ \nu \ge \nu_m$} \\
  \end{array}\right.\;,
\label{eq:fnu1}
\eeq
having defined $\nu_c\equiv\nu(\gamma_c)$ and $\nu_m\equiv\nu(\gamma_m)$.

On the other hand, if the condition $\gamma_c>\gamma_m$ is satisfied,
only electrons with $\gamma_e > \gamma_c$ can cool efficiently.  In
this situation, called {\em slow cooling regime}, the observed flux
varies according to the relation 
\beq 
 F_\nu  = F_{\nu,{\rm max}} \left\{
  \begin{array}{ll}
     (\nu/\nu_m)^{1/3}, & \hbox{$ \nu < \nu_m$} \\
     (\nu/\nu_m)^{-(p-1)/2}, & \hbox{$ \nu_m\le \nu < \nu_c$} \\
     (\nu_c/\nu_m)^{-(p-1)/2}(\nu/\nu_c)^{-p/2}, & \hbox{$ \nu \ge \nu_c$} \\
  \end{array}\right.\;.
\eeq 
In the above two equations, the parameter $p$ is the power-law index of the
electron energy distribution, while the quantity $F_{\nu,{\rm max}}$
represents the maximum value of the flux in the afterglow spectrum.
This is achieved when the observing frequency is $\nu=\nu_c$ in the
fast cooling regime, and when $\nu=\nu_m$ in the slow cooling regime.
Under the assumption that the magnetic field energy
density in the shell rest frame is a fraction $\xi_B$ of the
equipartition value, and that the power-law electrons carry a fraction
$\xi_e$ of the dissipated energy, this maximum intensity of the
afterglow flux is given by 
\beq
F_{\nu,{\rm max}}= 110\;n^{1/2}{\xi_B}^{1/2}\;E_{52}d_{28}^{-2}\;(1+z)\;{\rm mJy}\;,
\label{eq:Fnu}
\eeq
where $d_{28}$ is the luminosity distance in units of $10^{28}$ cm,
$n$ is the mean density of the surrounding medium in units of cm$^{-3}$,
and $t_d$ is the time in days, as measured in the observer frame, since 
the beginning of the burst. Here and in the following a 
fully adiabatic schock is assumed. Then
the cooling frequency $\nu_c$ and the 
synchrotron frequency $\nu_m$ are respectively given by (Sari et al. 1998)
\beq
\nu_c(t) = 2.7\times 10^{12}\;{n_1}^{-1} {\xi_B}^{-3/2} 
E_{52}^{-1/2}t_{\rm d}^{-1/2}\;(1+z)^{-1/2}\;{\rm Hz}\;,
\label{eq:nuc}
\eeq
and
\beq
\nu_m(t)=5.7\times 10^{14}\;{\xi_e}^2 {\xi_B}^{1/2} 
E_{52}^{1/2}t_{\rm d}^{-3/2}\;(1+z)^{1/2}\;{\rm Hz}\;.
\label{eq:num}
\eeq
The transition between the fast and the slow cooling regimes occurs at the time
defined by $\nu_c(t_0)=\nu_m(t_0)\equiv\nu_0$, 
\beq
t_0 = 210 \xi_B^2 \xi_e^2 E_{\rm 52} n_1\;\; {\rm days},
\label{eq:t0}
\eeq
and the corresponding frequency is
\beq
\nu_0 = 1.8\times10^{11}\xi_B^{-5/2}\xi_e^{-1}E_{52}^{-1}n_1^{-3/2} {\rm Hz}\;,
\label{eq:nu0}
\eeq
always under the assumption of adiabatic shock. When $\nu_{\rm obs}>\nu_0$, 
the flux reaches its peak value  at $t<t_0$, i.e. in the fast cooling regime, 
and therefore at the time when $\nu_c(t_c)=\nu_{\rm obs}$, that is
\beq
t_c = 7.3\times 10^{-6}\xi_B^{-3}E_{52}^{-1}n_1^{-2}\nu_{15,obs}^{-2} (1+z)^{-1}
\;\;{\rm days}\;.
\label{eq:tc}
\eeq
On the other hand, if $\nu_{\rm obs}<\nu_0$, the flux peakes during the slow
cooling regime, corresponding to the time at which $\nu_m(t_m)=\nu_{\rm obs}$,
where
\beq
t_m = 0.69\xi_B^{1/3}\xi_e^{4/3}E_{52}^{1/3}\nu_{15,obs}^{-2/3}(1+z)^{1/3}
\;\;{\rm days}.
\label{eq:tm}
\eeq 
The fraction of energy that goes into electrons, $\xi_e$, and the
fraction that is shared by the magnetic field, $\xi_B$, are not 
well-constrained either theoretically nor observationally, and they are
likely to vary significantly from burst to burst (see e.g. Kumar 1999). 
Therefore we draw them from a distribution
which we assume uniform in an interval that encompasses, for each of
them, the observationally inferred values in those few cases where 
they could be derived from fits to the light curve (Wijers \& Galama 1999;
Panaitescu \& Kumar 2001a) and the theoretical
expectations (e.g. Waxman 1997). More specifically, for each burst, we draw $\xi_e$ from
a uniform distribution in the interval (0.01,0.2), and $\xi_B$ from a
uniform distribution in the interval (0.001,0.1).

Finally, simulations of mergers between two compact objects have shown
that a typical energy release in $\gamma$ rays (deriving from
neutrino--antineutrino annihilation) is $E_\gamma\sim$ a few $\times
10^{50}$ ergs (Ruffert \& Janka 1999).  This is the isotropic
equivalent energy (i.e. after correcting for the typical beaming
angles found in the simulations) based on the neutrino annihilation
flux. However other channels (such as the energy of an accretion disk
that forms in a merger) are possible. Katz \& Canel (1996) report
$\sim 10^{51}$ ergs as typical energy in $\gamma$ rays\footnote{The fact
that the energy output of short bursts is smaller than that of their
longer counterparts is observationally established on the basis
of the fact that their peak fluxes are comparable to those of
long GRBs, while their durations are much smaller.}.  
Assuming an efficiency $\sim 0.2$
of conversion to $\gamma$ rays (Guetta, Spada \& Waxman 2001), we
adopt a typical value of $E=5\times 10^{51}$ ergs for the total isotropic
equivalent energy of the bursts. Small variations around this value
would not really affect the computed afterglow distributions, given
the spread in the other parameters. However the dependence of our
final results on the assumed value of the energy (cfr. Eqs
(\ref{eq:fnu1})-(\ref{eq:tc}) should be kept in mind.

In this work, we are primarily interested on some characteristic
quantities of the afterglow\footnote{When we talk about afterglow here
we always refer to the emission produced when the blast wave interacts
with the external medium, as discussed above. Kumar \& Panaitescu
(2000) have shown that afterglow emission in the X-ray can also be
produced by off-axis emission during the early times of the
GRB. However this emission (which could dominate in the first hour or
so if the density of the medium is very low) can be distinguished from
the traditional one based on the differences between their spectral and temporal
slopes.}  that would be the strongest diagnostics of a population of
mergers of compact objects in long-lived binaries, therefore those
quantities dependent on the location, and, in turn, on the density of
the medium. The peak flux scales with $n^{1/2}$, and therefore
provides a good diagnostic of the environment. A much stronger probe
of the density is the time $t_c$ at which the cooling frequency $\nu_c$ is
equal to the frequency of observation. This time depends on $n^{-2}$
and $E^{-1}$ [cfr. Eq. (\ref{eq:tc})], and therefore it is expected
that, given the reasonable assumption that the typical range of values
of $\xi_B$ in the afterglow is independent of the progenitor, the
distribution of these afterglow characteristic quantities should be
significantly different for the two classes of long and short bursts if they
are indeed associated with two distinct classes of progenitors, one
short-lived and the other long-lived\footnote{Also note that the probably lower
energy of the mergers with respect to the collapsars 
would tend to make the two distributions even
more different.}.

\section{Characteristics of a population of GRBs due to mergers}

The computation of the properties of this population involves several
steps, all of which are Monte Carlo based. First, a random redshift is
generated from the rate in Eq. (\ref{eq:Rmerg}).  At that redshift,
the mass of the host galaxy is randomly drawn from the distribution in 
Eq.(\ref{eq:pgal}), and a random inclination with respect to the plane
perpendicular to the line of sight to the observer is assigned to it.
The location of the burst in terms of the coordinates ($R, \eta$) is
then randomly drawn from the probability distribution obtained with
the population synthesis code for the galaxy of that given mass at
that given redsift.  The density of the medium at the location
($R,\eta$) for the host galaxy of mass $M_{\rm gal}$ is hence found
from Equations (\ref{eq:rhoh}) and (\ref{eq:rhobd}), and finally the
afterglow properties of that given burst are computed as described in
\S 2.4.

The results of the simulations for a randomly generated sample of
10,000 merger events (for each of the two progenitor types NS-NS and
NS-BH) are displayed in Figures 2-7.  All these results have been derived  
for a model with galaxy masses in the range $\{10^8-10^{11} M\odot\}$ and
with parameter $\beta=1$ (i.e. merger probability in a galaxy of
a given mass directly proportional to the galaxy mass).  
We show both the differential distributions (which give a better visual idea
of where most of the events are) and the cumulative distributions, which
better quantify our results. 
To start with, in  Figure 2 we show the distances ($d_{\rm proj}$) from the galaxy
centers projected onto the plane perpendicular to the line of sight to
the observer.  Given their rather short lifetime, NS-NS events occur
rather close to the galactic centers. Fig. 3 shows the distribution of
physical offsets $\theta=d_{\rm proj}/d_A$ (where $d_A$ is the angular
diameter distance) corresponding to the projected distances of Fig. 2.

The typical ambient densities in which the events occur are shown in
Figure 4. NS-NS mergers probe typical ISM densites, while a
substantial fraction of NS-BH events occurs in a very low-density
environment. The values of the densities expected from NS-NS events
are consistent with those inferred from the afterglow modelling of
some of the long bursts. However, if the typical energies released in
mergers of compact objects (Ruffert \& Janka 1999) are indeed smaller
than those inferred so far for long GRBs\footnote{For this class of
bursts Frail et al. find a ``typical'' energy of $3\times 10^{51}$
ergs after correcting for beaming effects.}  then the afterglows from
NS-NS mergers would be correspondingly dimmer. Fig. 5 shows the
expected peak in the afterglow spectrum (cfr. Eq. \ref{eq:Fnu}) with
the assumed isotropic energy release of $5\times 10^{51}$ ergs.

The rather larger distances from the galactic centers, at which NS-BH
mergers occur, result in significantly smaller typical densities of
the surrounding medium, as shown in Fig. 4. This distribution does not
appear consistent with densities inferred so far for the class of long GRBs.
The corresponding peak afterglow flux for the NS-BH events
(displayed in Fig. 5) is correspondingly rather lower ($F_{\rm
peak}\propto n^{0.5}$).  Here the same value of the  energy of
$5\times 10^{51}$ ergs
has been adopted, while, as explained in \S 2.4, the other afterglow parameters
have been drawn from distributions consistent with values inferred for some
of the long bursts. 

It should be noted that the distributions of $F_{\rm peak}$ should be
considered mostly representative in the X-ray band.  Predictions
in the longer wavelength bands are affected by other factors that we
have not included here. In the optical, obscuration by dust is
considered the likely responsible for the failure to detect counterparts in
about half of the X-ray identifications of the long GRBs
(Djorgovski et al. 2001; Reichart 2001; Venemans \& Blain 2001; but
see Lazzati, Covino \& Ghisellini 2000b), and its
effect on the number counts have been shown to be quite sensitive to
the type of dust model considered (Perna \& Aguirre 2000). Moreover,
dust destruction by the X-ray UV photons (Waxman \& Draine 2000;
Reichart 2001; Draine \& Ho 2001) can introduce time variability in
the extinction curve and consequentely in the measured value of the
flux at various times. If the emission is beamed, a further reduction
in the flux is expected at later times (Sari, Piran \& Halpern 2001), 
but the frequency at which this happens depends on the specific value of the beaming angle.
For the particular value of the 
 density $n=10^{-3}$ cm$^{-3}$, and an energy on the order of
$5\times 10^{51}$ erg, Panaitescu, Kumar \& Narayan (2001) note that optical
and radio afterglows of short bursts are likely to be below current
detection limits even without accounting for all the other effects.

Figure 6 shows, for both the case of NS-NS mergers (Fig.6a) and NS-BH
mergers (Fig.6b), the integrated afterglow flux in the 2-10 keV energy
band at the observation times $t_{\rm obs}=1$hr, 3hr, 12hr after the
burst.  
Given the
detection thresholds of the current X-ray instruments {\em SAX} and
{\em CXO}, and the upcoming {\em Swift} (in the range of a few $\times
\{10^{-15}-10^{-14}\}$ erg/cm$^2$/s), a sizeable fraction of events
should be detectable if observed within the first few hours.  These
simulations have been made with the choice $p=2$ for the power-law
index of the electron energy distribution, in agreement with the
average value inferred from numerical modelling of the afterglow by
Panaitescu \& Kumar (2001b). 
However, we want to emphasize the fact that the integrated
afterglow flux in the band we considered is rather sensitive to this
choice. We performed our simulations also with the value $p=2.5$
(while all the rest remaning the same), and found that the fluxes of
the bursts were typically reduced by a factor $\sim 10-20$.  In
general, it is reasonable to expect a distribution of values of $p$
among bursts, but most likely not as wide as that of other parameters.

The inference of the physical parameters (such as $E$, $\xi_B$,
$\xi_e$, $n$) which characterize a burst and its afterglow, from an
analysis of the light curve, requires the observational determination
of the characteristics breaks at $\nu_c$, $\nu_m$, $\nu_a$ (absorption
frequency) and a measurement of the peak flux (see e.g. Wijers \& Galama
1999). This in turn requires,
for each burst, a coverage over a wide range of frequencies. On the
other hand, some statistical information can be obtained by measuring
a smaller subset of these quantities, but for a larger sample of
bursts.  A very good probe of the density of the medium is the time
$t_c$ at which the cooling frequency $\nu_c$ is equal to the
observation frequency $\nu_{\rm obs}$. This time, in fact, depends on
$n^{-2}$ [cfr. Eq. (\ref{eq:tc})]. Even with a rather large spread in
the values of $\xi_B$, the sensitivity to the density remains
very strong. The distribution of cooling times can be particularly
useful in identifying differences in the environment between the populations
of long and short bursts. In fact, assuming as reasonable that the
physics of the afterglow is the same for the two populations, and therefore
that the distribution of the parameter $\xi_B$ is also the same, then
the distributions of cooling times should significantly differ for
the populations of long and short bursts if their progenitors occur in
different environments. Fig. 7 shows the distribution of $t_c$ (assuming
an observing frequency corresponding to 1 keV) for the NS-NS merger
scenario, and for the NS-BH case.  Note that times which are too short to allow
observation at the frequency considered here are longer at lower
frequencies ($t_c\propto\nu_{\rm obs}^{-2}$). They can therefore be measured
at longer wavelength and then simply rescaled to one given frequency to build
up the distribution.  

While the above simulations of the events were computed for a model
with merger rates proportional to the galaxy mass, we explored how
our results changed when a relatively higher weight for the rates
was given to galaxies with smaller masses (i.e. the model with $\beta=0.5$
in Eq.{\ref{eq:pgal}). The mass distribution obtained with this
model is compared with the other one ($\beta=1$) in Fig. 8, for both 
populations of NS-NS mergers and NS-BH mergers.  
Because of their shorter lifetimes, NS-NS mergers typically occur 
at higher reshifts than NS-BH mergers, and therefore they have a
higher probability of occurring in small mass galaxies than the NS-BH
group. This is the reason for the behaviour of the two populations shown 
in Fig. 8. 

The results of the simulations with the two different mass
distributions of Fig.8 are shown and compared in Fig.9 for the
distribution of projected distances. When the typical galaxy mass is
reduced, there are two competing effects: first, at the beginning of
its evolution, the population is much closer to the galaxy center
due to the smaller size of the disks and therefore it has to travel
for a longer time to reach a given $d_{\rm proj}$; 
second, the reduced potential of the smaller-mass galaxies makes it
easier for the population to move further away from the center.  
As Fig. 9 shows, the first effect tends to dominate for the NS-NS
population, due to its very short lifetime. If more of these events
occur in smaller galaxies, then the offsets from the galaxy centers
will also be typically  smaller. On the other hand, for the population
of NS-BH mergers, which has a longer lifetime, the second effect prevails,
and an enhanced merger rate in small-mass galaxies results in generally
larger distances from the galactic centers.

\section{Discussion and conclusions}

We have studied the properties that a population of GRB events due to
mergers of compact objects should have with special emphasis on the
related afterglows.  By using a Monte Carlo type of
approach, our simulations take into account the mass distribution of
the host galaxies as a function of redshift, as well as the redshift
evolution of the probability distributions for the location of the
mergers within galaxies of various mass (cfr. Fig. 1abc). This last 
effect
needs to be taken into account especially for binaries whose merger time
can be comparable with the Hubble time for a sizeable fraction of them
(such as the NS-BH population).

Our population of double neutron star binaries includes the new groups
of short-lived binaries identified by Belczynski \& Kalogera (2001) and 
Belczynski et al.\ (2001a, 2001c), which dominate the 
merger rates. Therefore our results regarding this population
differ from previous studies on the same subject, and the derived
distributions trace rather closely the star forming regions in the disk.
The densities in which they occur are typical ISM densities; hence
their afterglows, even though dimmer due to the smaller energy
released, should still be observable with current X-ray instruments for
a large fraction of them. Afterglows
produced as a result of NS-BH mergers are even dimmer, but most of them should
still be detectable if observed within the first few hours.   

Whereas the NS-NS class of candidate GRB progenitors 
might not be distinguishible from that of collapsars
and of other promptly-bursting binaries simply on the basis of their
location within the host (and the consequent intensity of their
afterglows), there are however other signatures, such as the 
presence of an underlying supernova explosion, or of iron lines
in the afterglow spectrum (see Introduction) that, while naturally associated 
with a collapsar, would be hard to explain within the NS-NS merger
scenario\footnote{It should however be noted that ``SN bumps'' can
also be explained as the result of dust echos (Esin \& Blandford
2000), and that models that explain iron lines (e.g. Vietri \& Stella
1998) require a SN that took place a few months before the GRB, which
is inconsistent with the presence of the SN bumps.}. 

On the other hand, the observational properties of the NS-BH binary population
that we have studied here, and which depend on the location within the hosts
(such as offsets, densities, and density-dependent afterglow quantities),
differ significantly from those of the NS-NS population (which, as far
as the location is concerned, could well be representative also of collapsars).
Even if it is not possible to infer all the parameters $E\;,n,\;\xi_e,\;\xi_B$
at once (which requires that $\nu_c$, $\nu_m$, $\nu_a$ and $F_{\rm peak}$
be all measured), a {\em comparison} of the distributions for, say, $F_{\rm peak}$
or $t_c$ for the class of long and that of short bursts would provide strong
constraints on whether their progenitors actually belong to two different classes
of progenitors, one which is short-lived and the other which is long-lived.

The population synthesis code that we used in all the simulations is
the {\em StarTrack} code by Belczynski et al. (2001c), operated in its
``standard'' mode, where the best values of all the parameters are
chosen.  A complete parametric study of how the distributions for the
location within a galaxy of a given mass change when all the model
parameters are varied to their extremes is being performed elsewhere
(Belczynski et al.\ 2001b). The main results are that the merger site
distribution has the strongest dependence on the prescriptions for the
mass transfer and the common envelope efficiency, and it is also
rather dependent on the maximum allowed NS mass and the kick velocity.
It is not very dependent on the assumed cosmology. Regarding the
new class of short-lived binaries identified by Belczynski \& Kalogera
(2001) and Belczynski et al. (2001a, 2001c), it is found that 81\% of 
them contribute to the NS-NS population in our standard model described 
in \S 2.1. The parametric study shows that  the highest contribution (98\%) 
from this population is obtained for small kicks, while the smallest
(28\%) for very low CE efficiency. We stress once again that our
results on GRBs from NS-NS mergers strongly rely
on the presence of this short-lived population, whose presence is based
on the assumption that low-mass helium stars can survive the CE phase.
This will have to be tested through detailed hydrodynamical
simulations.  In this work, 
our main interest  has been to incorporate the
results of the population synthesis code within a proper cosmological
context, and study the expected afterglows if GRBs are indeed
associated with mergers of two compact objects. This is particularly
relevant for the population of short bursts, whose very short
timescales are hard to account for with the collapse of a massive
star, while being naturally associated with mergers of two compact
objects.

\acknowledgements We want to thank several people for very useful
discussions on various aspects of this project. In particular, we are
indebted to Tomasz Bulik, Vicky Kalogera, Chris Kochanek, Davide
Lazzati and Ramesh Narayan. RP acknowledges support by the Harvard
Society of Fellows, and by the Harvard-Smithsonian center for
Astrophysics for a research grant. KB acknowledges support by the
Smithsonian Institution through a CfA Predoctoral Fellowship and by
the Polish Nat.\ Res.\ Comm.\ (KBN) grants 2P03D02219, 5P03D01120.  KB
also acknowledges support by the Polish Science Fundation (FNP)
through a 2001 Polish Young Scientist Award.

\newpage

{Fig.1 --- Distribution of projected distances (onto the observer plane)
for a face-on galaxy at different redshifts
(here $d_{\rm proj}=R$, coordinate in the plane of the disk).
Both populations of NS-NS and NS-BH mergers are considered,
and the mass of the galaxy is $10^8 M_\odot$ in Fig1a, 
$6\times 10^{9} M_\odot$ in Fig. 1b, and 
$10^{11} M_\odot$ in Fig.1c.}

{Fig.2 --- Distribution of projected distances for the two populations
of NS-NS mergers and NS-BH mergers. The results here (and up to Fig.7) 
are for a simulation with host galaxy masses in the range 
$10^8-10^{11} M_\odot$ and merger
rates proportional to the galactic masses (i.e. $\beta=1$).}

{Fig.3 --- Offset distribution corresponding to the projected distance
distribution in Fig.2.}

{Fig.4 --- Density distribution for the environment in which the two
populations of NS-NS mergers and NS-BH mergers are expected to occur.}

{Fig.5 --- Peak flux in the afterglow spectrum expected for the
density distribution in Fig.\ref{fig:den} and afterglow parameters
in a range typically inferred for the afterglows of long GRBs.}

{Fig.6 --- Distribution of the integrated afterglow flux in the 2-10 keV
energy band at several observation times after the burst for 
both populations of NS-NS mergers (Fig.6a) and NH-BH mergers (Fig.6b). 
At early times, a considerable fraction of 
events should be detectable with current X-ray instruments.}

{Fig.7 --- Times in the observer frame at which the cooling frequency 
is equal to an observation frequency corresponding to 1 keV. These
times are very sensitive to the density of the medium ($t_c\propto n^{-2}$).}

{Fig.8 --- Mass distribution for the two models considered here: one where
the merger rates are assumed proportional to the galaxy mass ($\beta=1$),
and the other which accounts for a possible rate increase in small mass
galaxies ($\beta=0.5$).}

{Fig.9 --- Comparison between the distributions of projected distances 
obtained for a model where the merger rates are simply proportional to 
the galaxy mass, and  a model which accounts for a possible rate increase in small mass
galaxies (the mass distributions in the two models are shown in Fig.8).
Larger $d_{\rm proj}$ correlate with larger galaxy masses for the NS-NS
population, while they generally anticorrelate for the NS-BH population whose lifetime
is a sizeable fraction of the Hubble time.} 

\newpage

\newcounter{figmain}
\newcounter{figsub}[figmain]
\renewcommand{\thefigure}{\arabic{figmain}.\alph{figsub}}

\refstepcounter{figmain}  

\refstepcounter{figsub}
\begin{figure}[t]
\centerline{\epsfysize=5.7in\epsffile{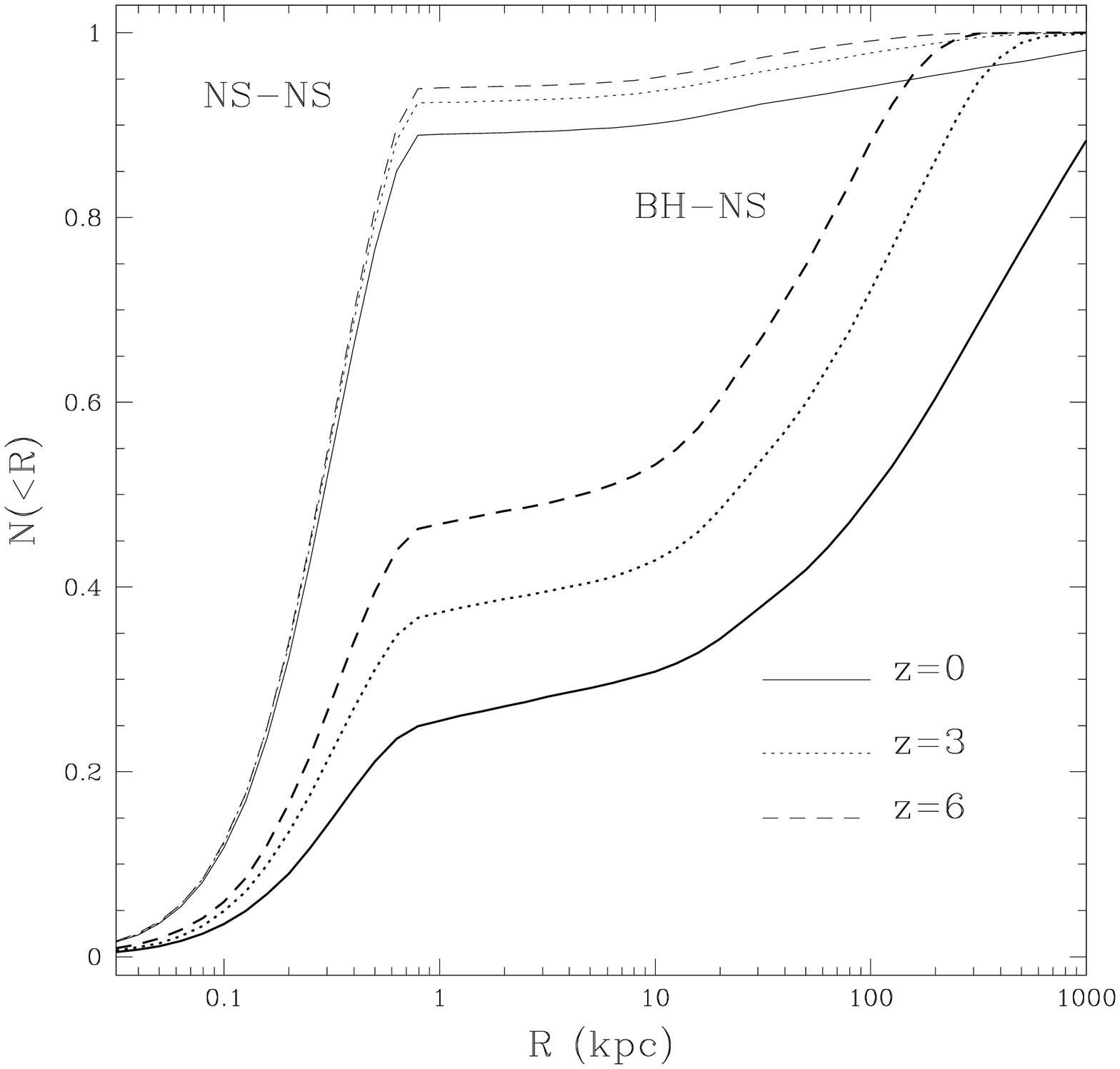}}
\label{fig:evol1} 
\end{figure} 

\refstepcounter{figsub}
\begin{figure}[t]
\centerline{\epsfysize=5.7in\epsffile{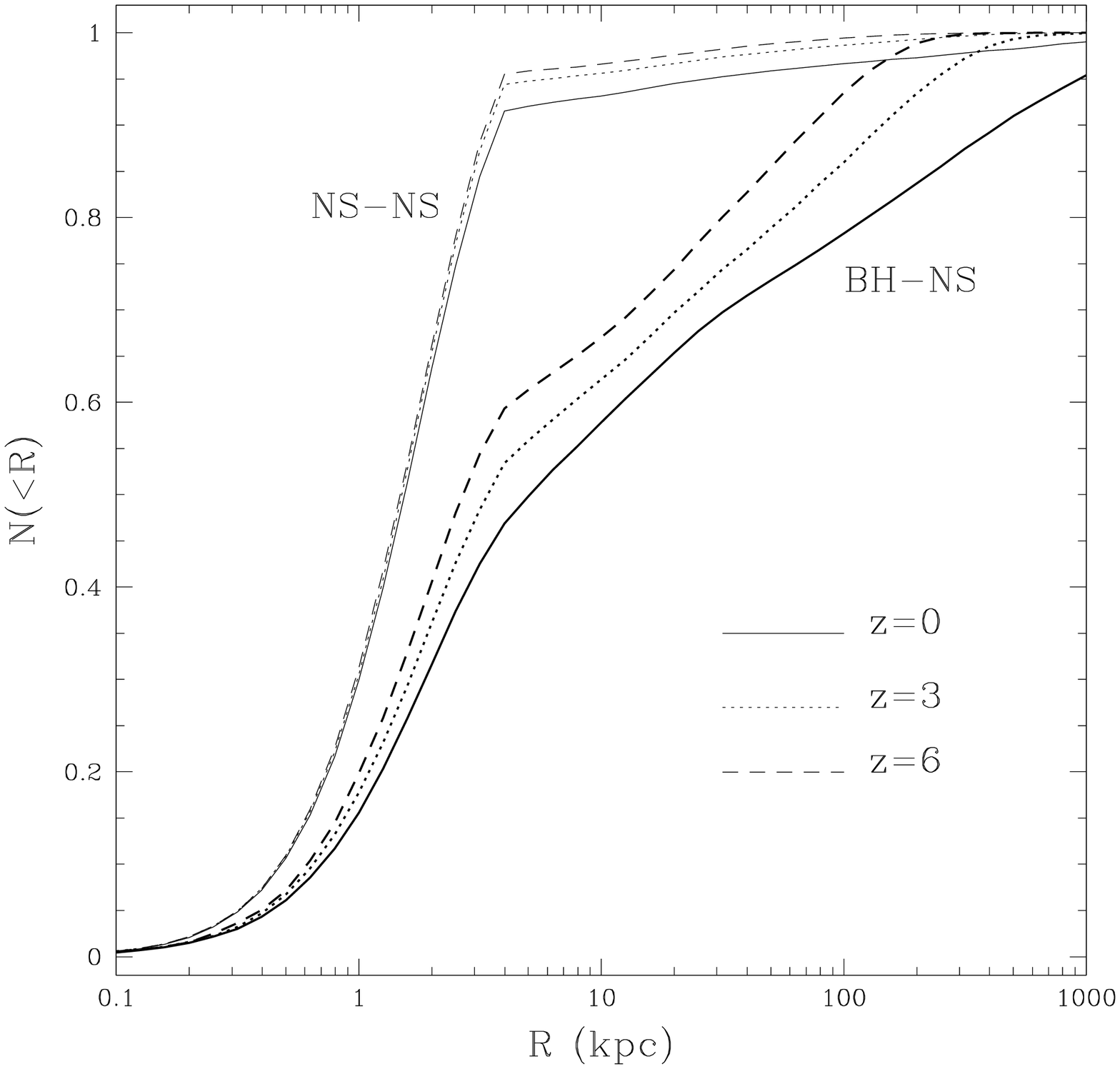}}
\label{fig:evol2} 
\end{figure} 

\refstepcounter{figsub}
\begin{figure}[t]
\centerline{\epsfysize=5.7in\epsffile{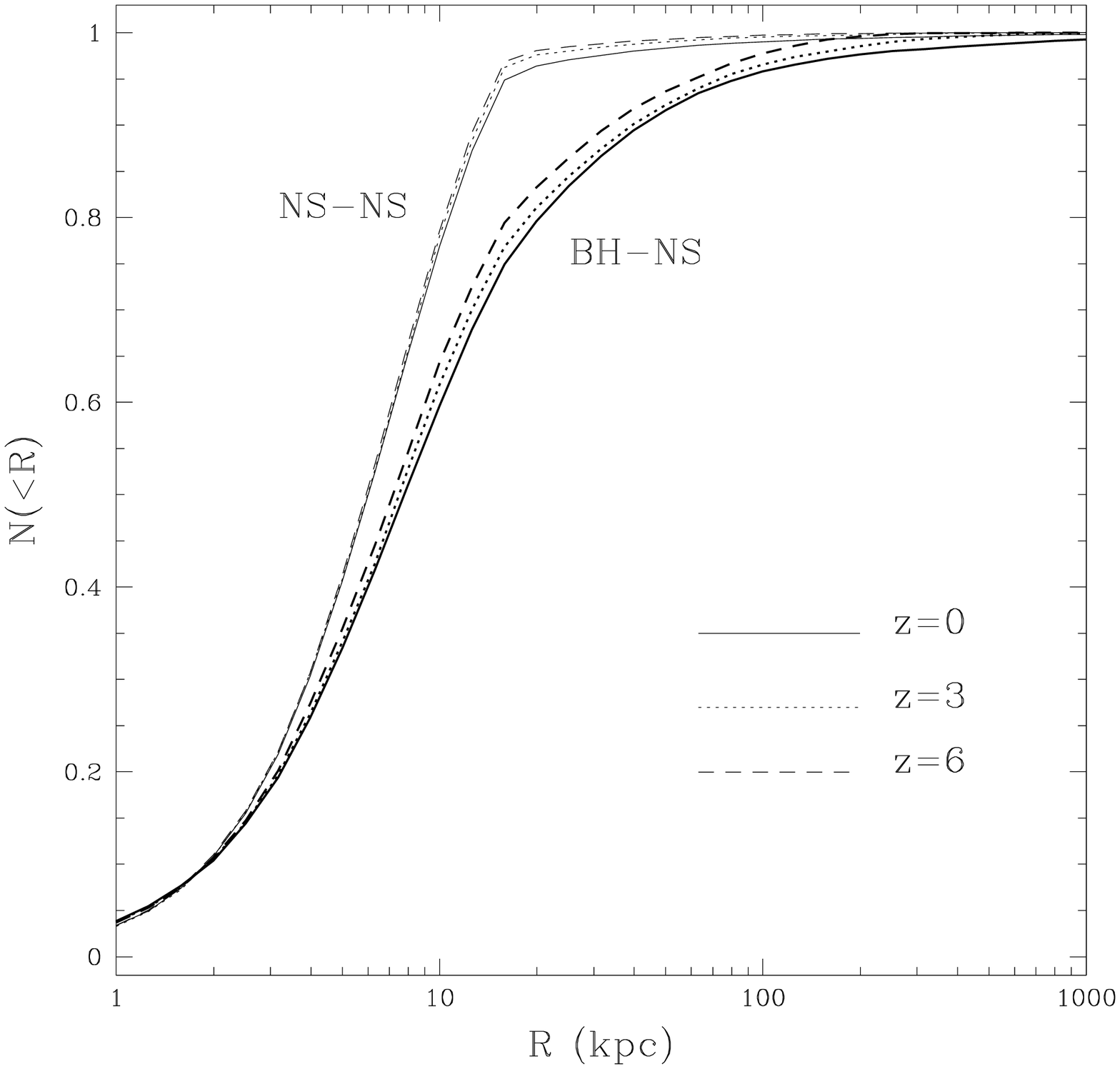}}
\label{fig:evol3} 
\end{figure} 

\refstepcounter{figmain}
\begin{figure}[t]
\centerline{\epsfysize=5.7in\epsffile{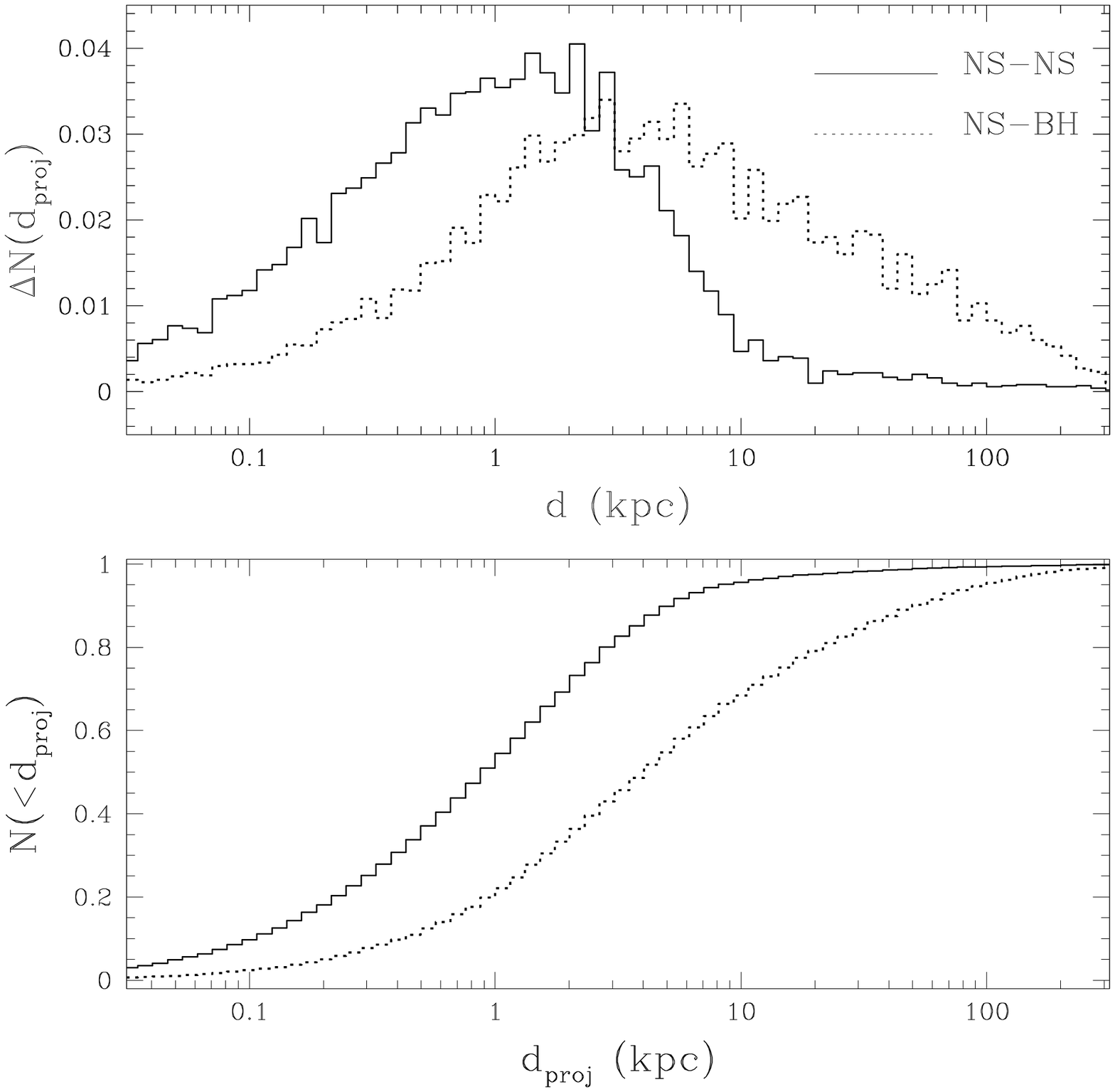}}
\label{fig:dproj} 
\end{figure} 

\refstepcounter{figmain}
\begin{figure}[t]
\centerline{\epsfysize=5.7in\epsffile{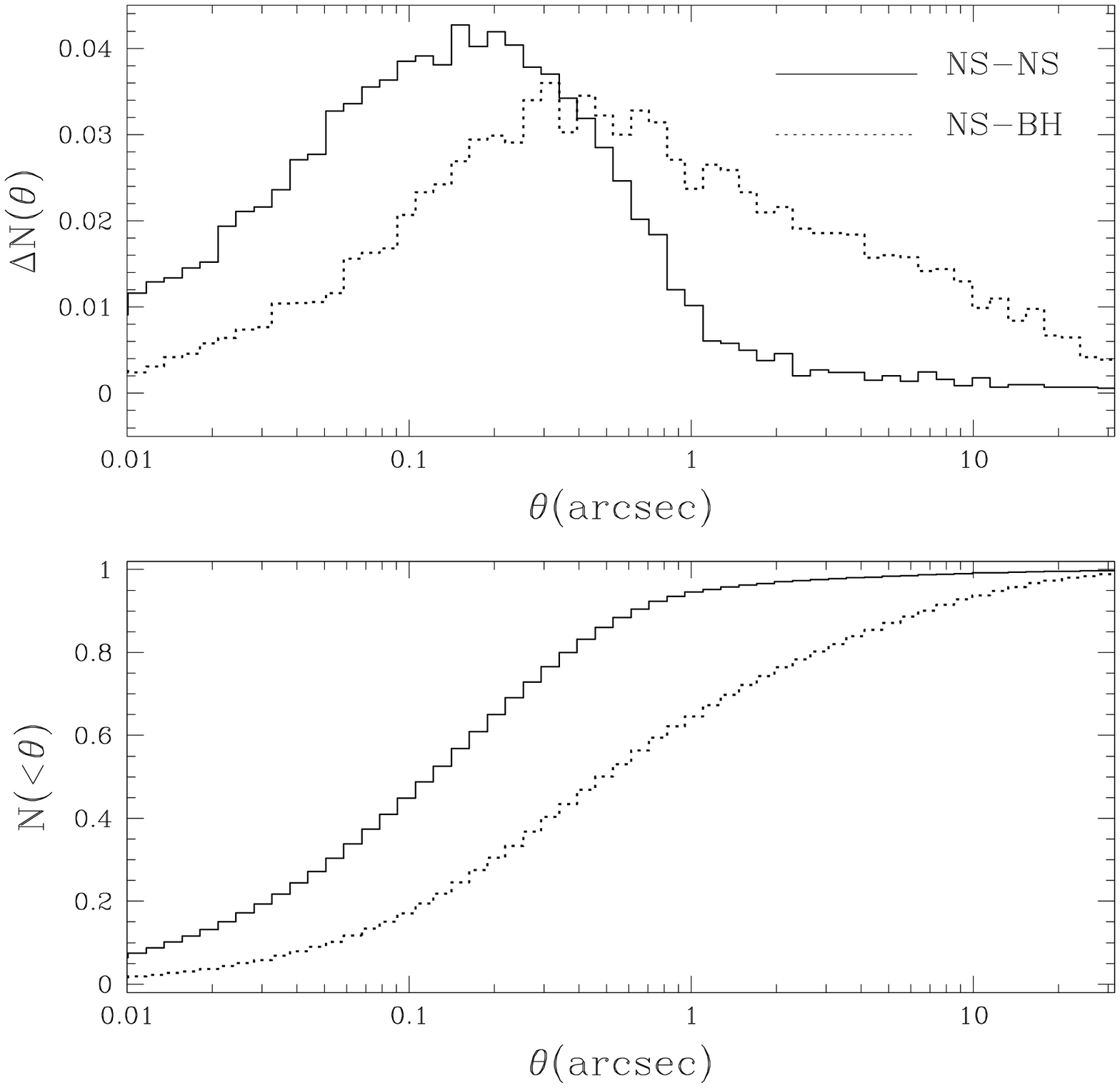}}
\label{fig:offset} 
\end{figure} 

\refstepcounter{figmain}
\begin{figure}[t]
\centerline{\epsfysize=5.7in\epsffile{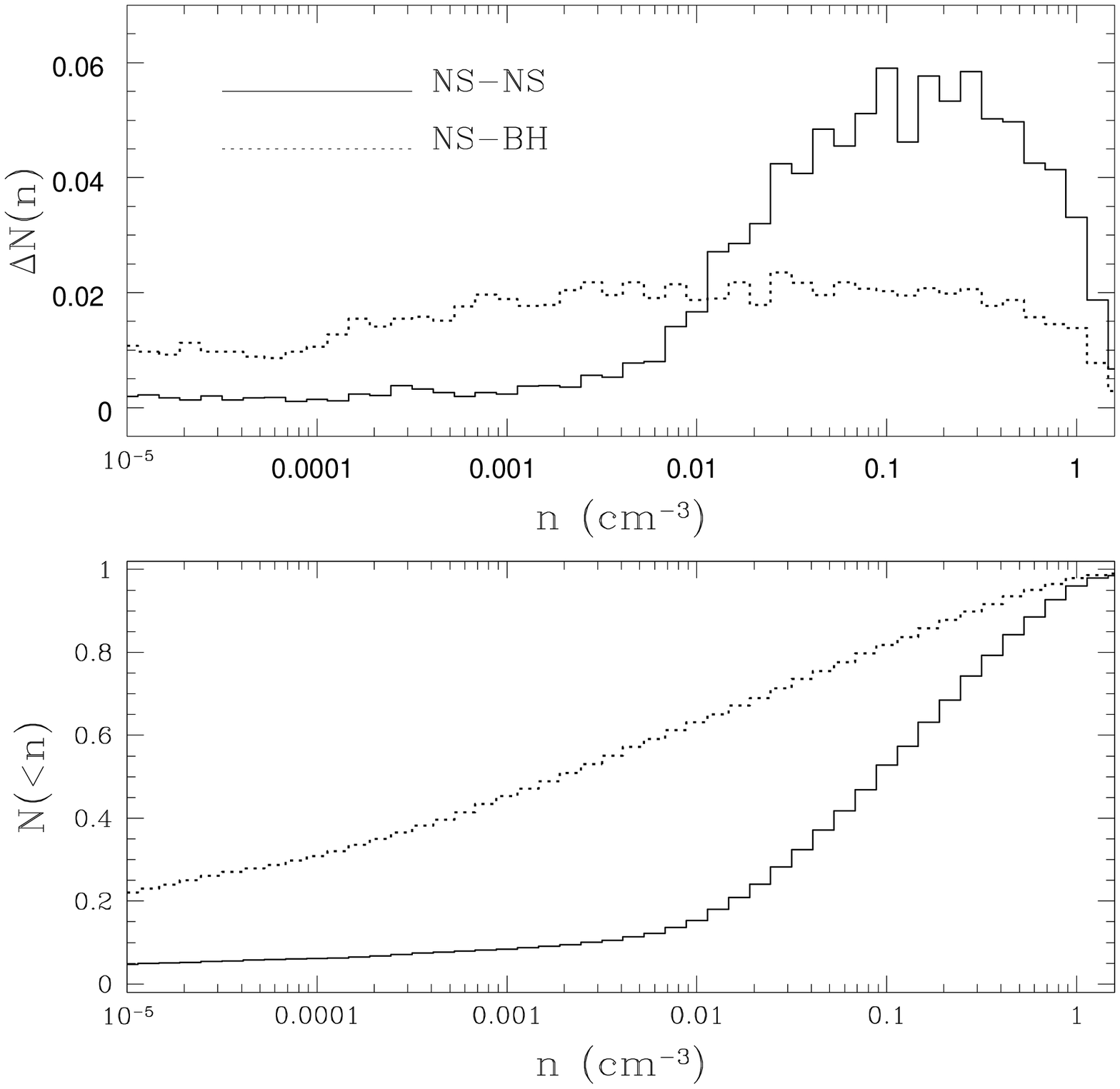}}
\label{fig:den} 
\end{figure} 

\refstepcounter{figmain}
\begin{figure}[t]
\centerline{\epsfysize=5.7in\epsffile{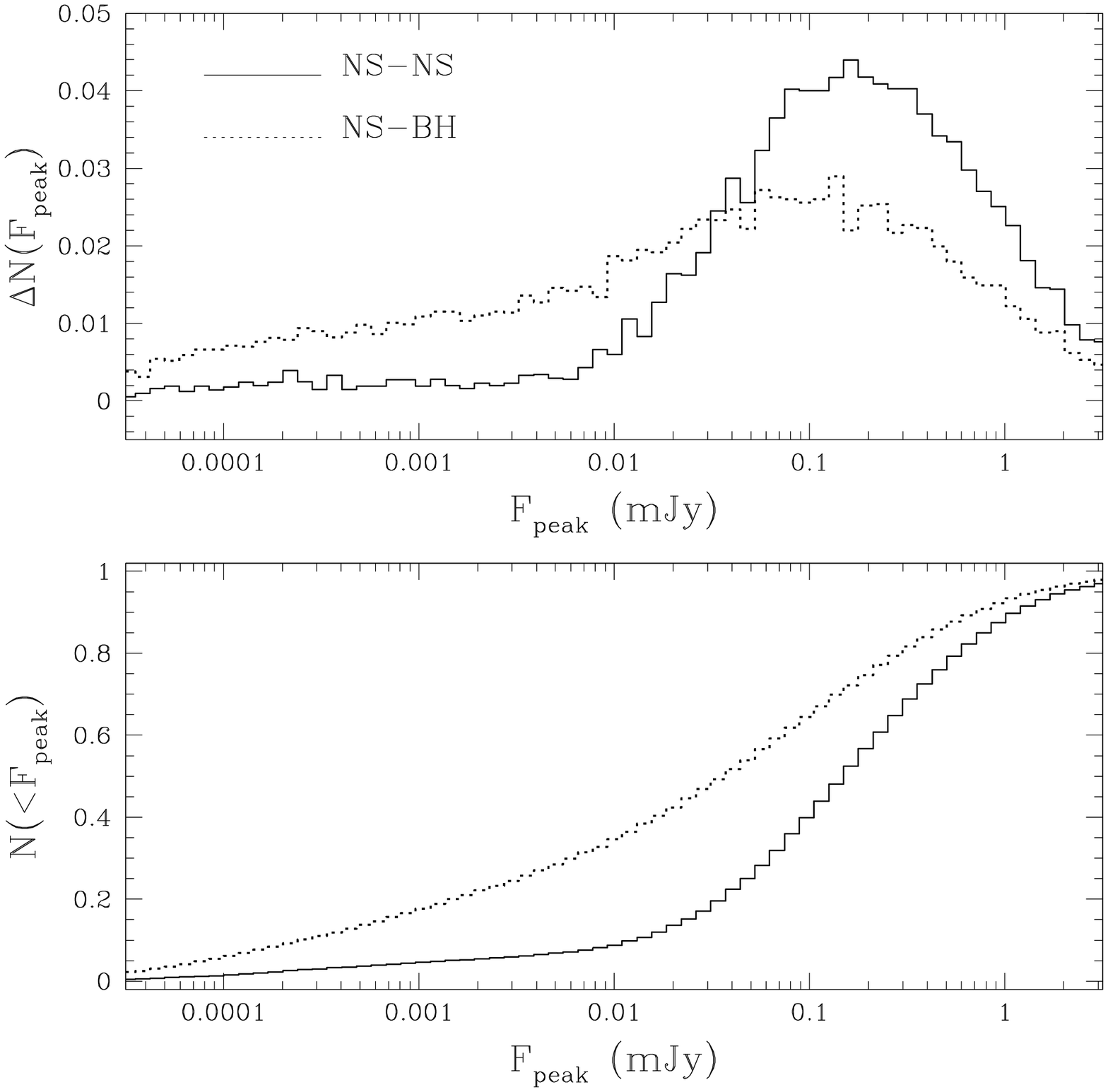}}
\label{fig:Fpeak} 
\end{figure} 

\refstepcounter{figmain}
\refstepcounter{figsub}
\begin{figure}[t]
\centerline{\epsfysize=5.7in\epsffile{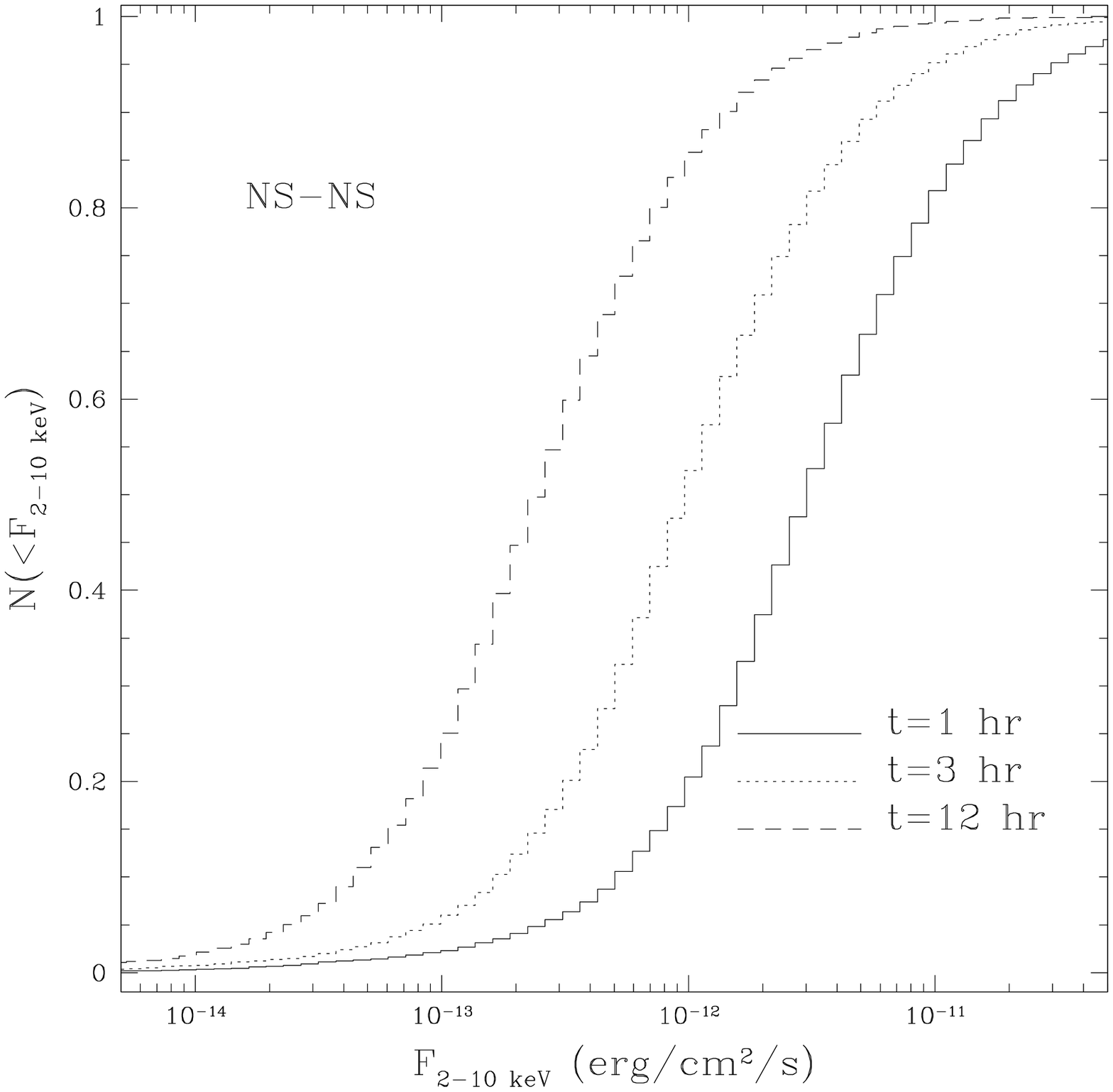}}
\label{fig:F2_10ns} 
\end{figure} 

\refstepcounter{figsub}
\begin{figure}[t]
\centerline{\epsfysize=5.7in\epsffile{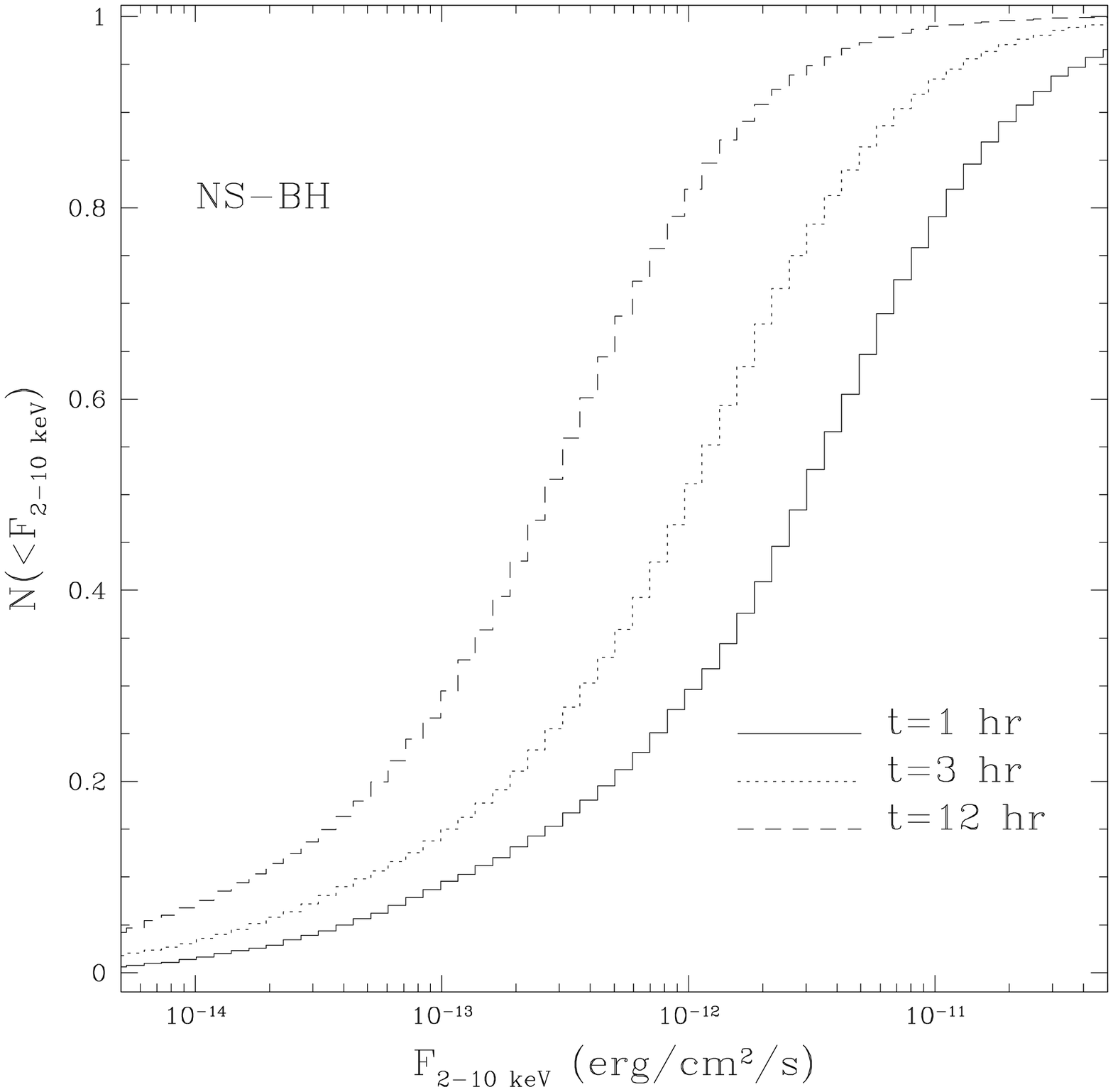}}
\label{fig:F2_10bh} 
\end{figure}

\refstepcounter{figmain}
\begin{figure}[t]
\centerline{\epsfysize=5.7in\epsffile{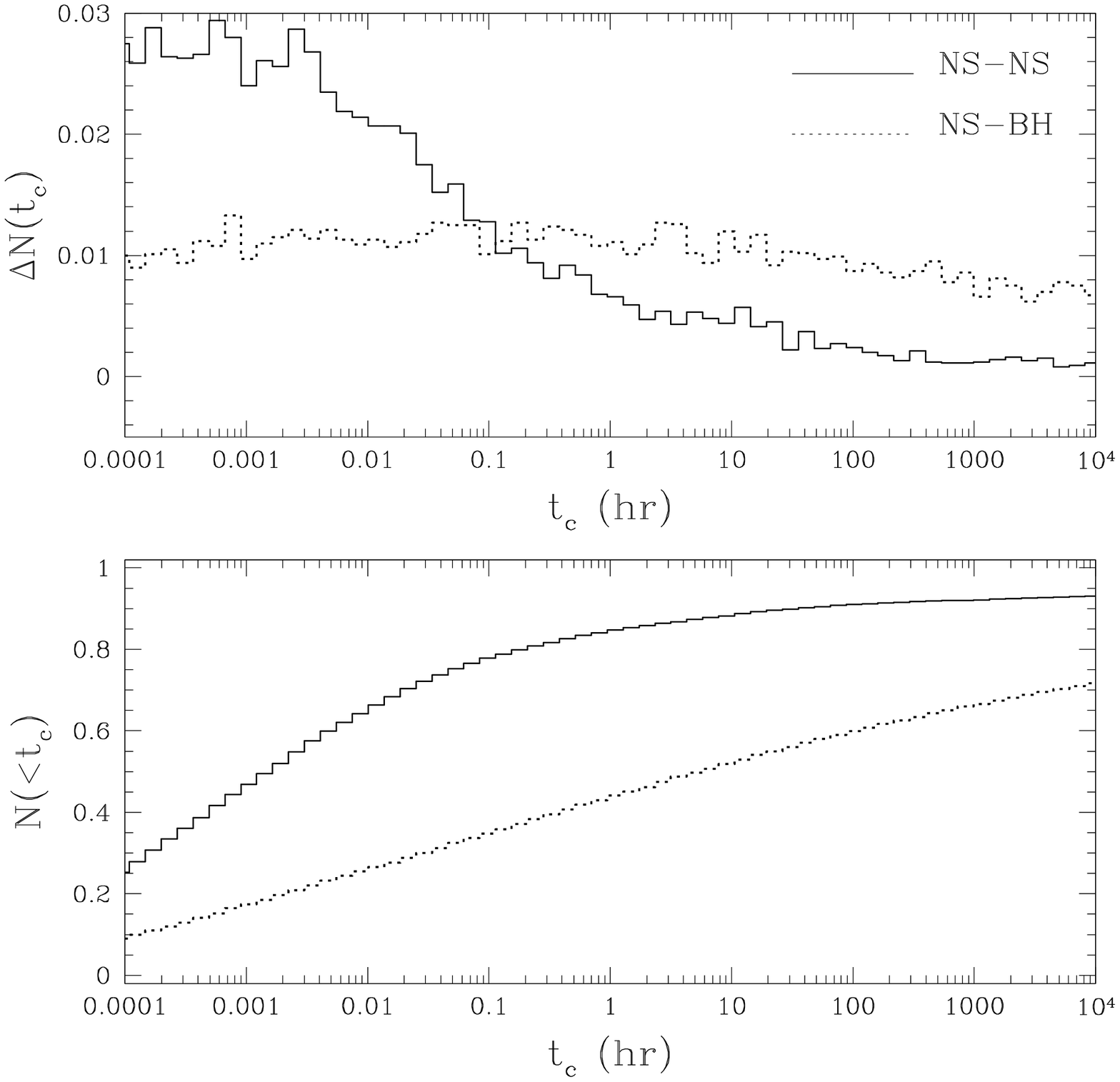}}
\label{fig:tc} 
\end{figure}

\refstepcounter{figmain}
\begin{figure}[t]
\centerline{\epsfysize=5.7in\epsffile{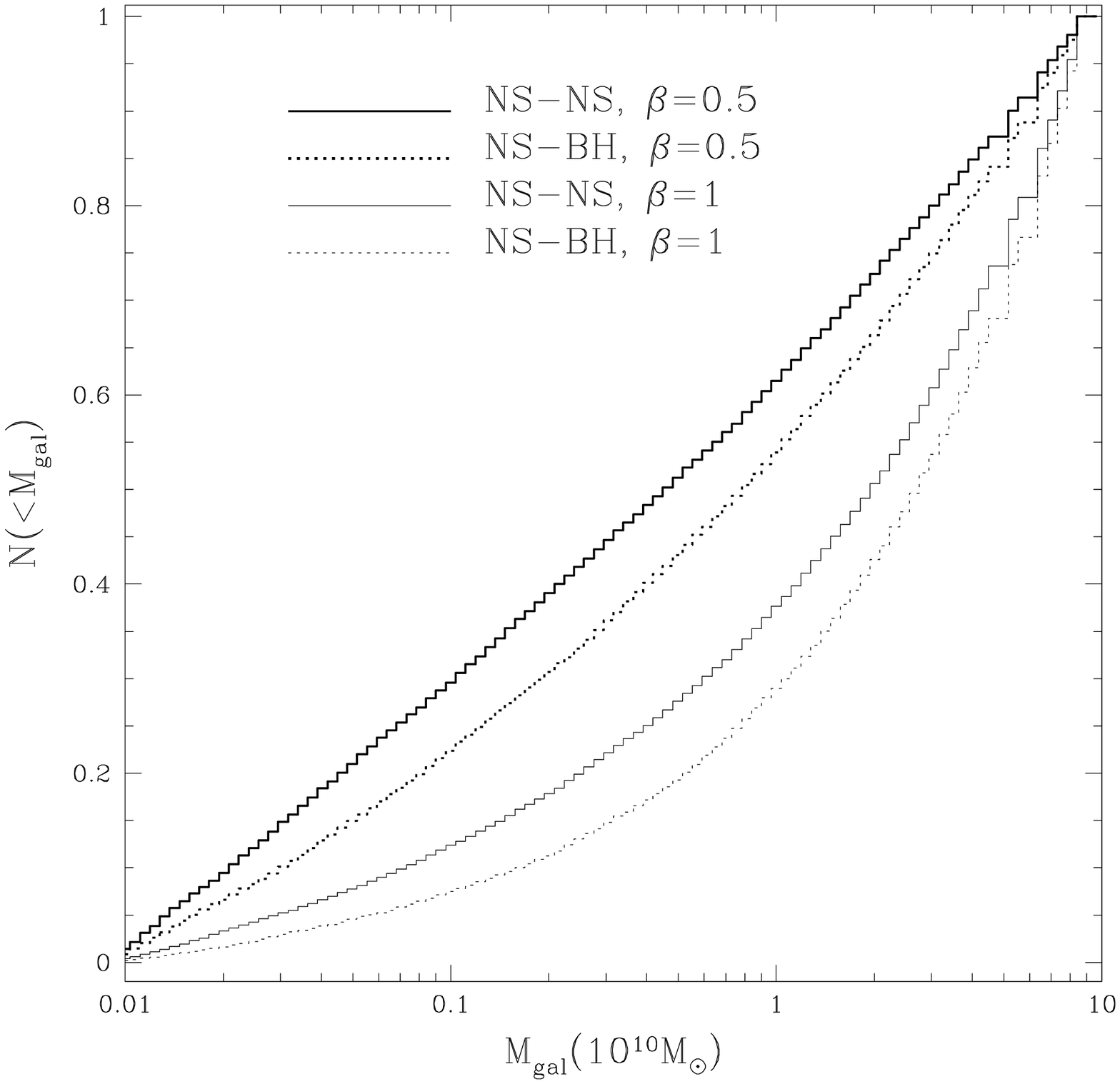}}
\label{fig:gal2} 
\end{figure}

\refstepcounter{figmain}
\begin{figure}[t]
\centerline{\epsfysize=5.7in\epsffile{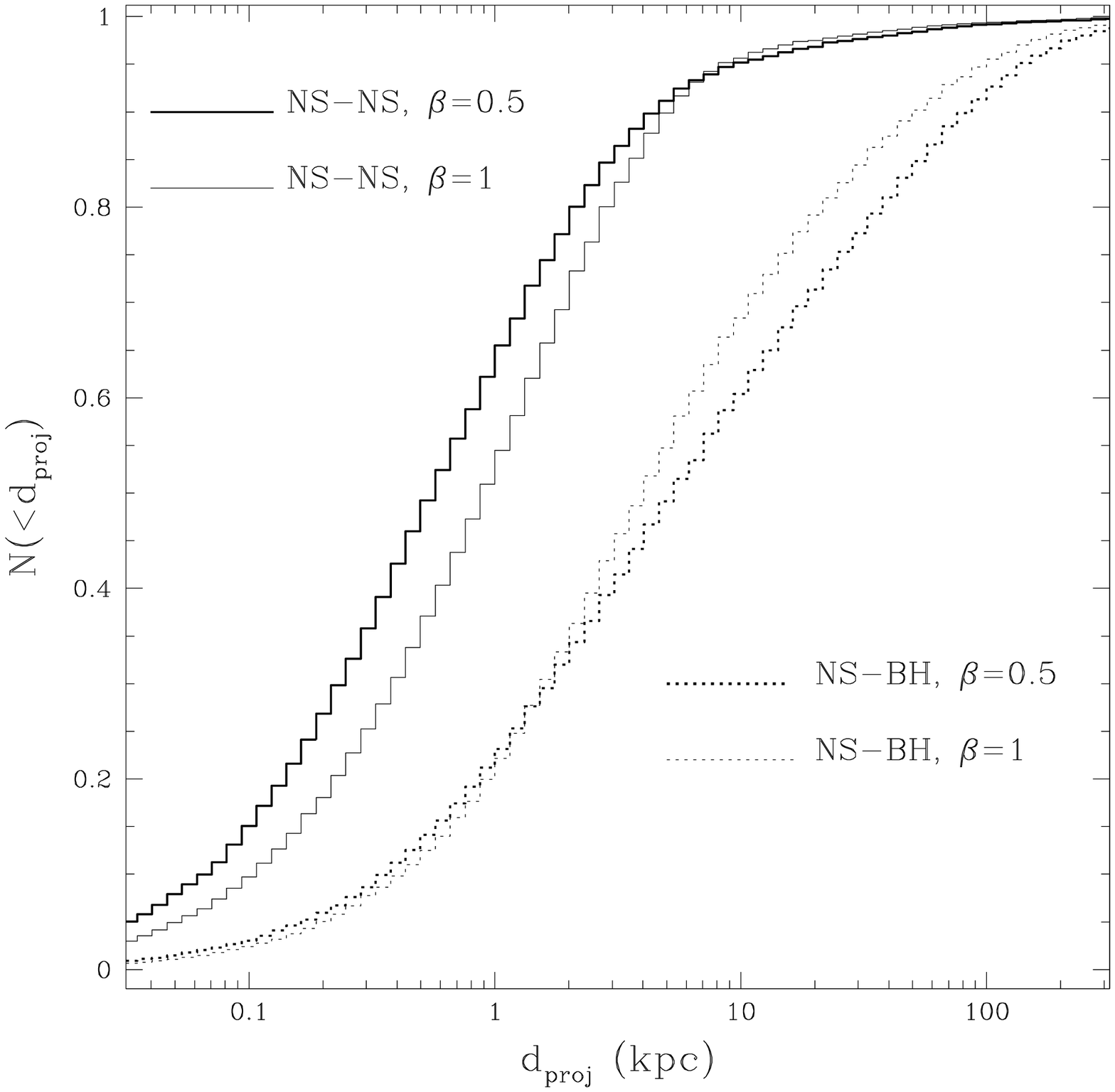}}
\label{fig:dproj2} 
\end{figure}

\end{document}